\documentclass[11pt,fleqn]{article}

\usepackage{amsmath,amssymb,amsthm,enumerate,cite,color}

\newcommand{\vspacebefore}{\raisebox{0ex}[2.5ex][0ex]{\null}}
\newcommand{\p}{\partial}
\newcommand{\const}{\mathop{\rm const}\nolimits}

\newcommand{\thetbn}{\arabic{nomer}}

\newcommand{\CV}{\mathop{\rm CV}\nolimits}
\newcommand{\CL}{\mathop{\rm CL}\nolimits}

\newcounter{tbn}
\newcounter{tabu}

\newcounter{mcasenum}

\newtheorem{theorem}{Theorem}

{\theoremstyle{definition} \newtheorem{definition}{Definition}

\newtheorem{note}{Note}
\newtheorem*{note*}{Note}
\begin{document}

\par\noindent {\LARGE\bf
Solutions and reductions for radiative energy\\ transport in laser-heated plasma.
\par}
{\vspace{4mm}\par\noindent {\bf P. Broadbridge~$^\dag$ and N.M. Ivanova~$^\ddag$
} \par\vspace{2mm}\par}
{\vspace{2mm}\par\noindent {\it
$^\dag{}$~School of Engineering $\&$ Mathematical Sciences, La Trobe University, Victoria 3086, Australia
\\
}}
{\noindent \vspace{2mm}{\it
$\phantom{^\dag}$~e-mail: P.Broadbridge@latrobe.edu.au
}\par}

{\vspace{2mm}\par\noindent {\it
$^\ddag$~Institute of Mathematics of NAS of Ukraine,~3 Tereshchenkivska Str., 01601 Kyiv, Ukraine\\
}}
{\noindent \vspace{2mm}{\it
$\phantom{^\ddag}$~e-mail: ivanova@imath.kiev.ua
}\par}

{\vspace{7mm}\par\noindent\hspace*{8mm}\parbox{140mm}{\small
Abstract\\

A full symmetry classification is given for models of energy transport in radiant plasma when the mass density is spatially variable and the diffusivity is nonlinear.
A systematic search for conservation laws also leads to some potential symmetries, and to an integrable nonlinear model.  Classical point symmetries, potential symmetries and nonclassical symmetries are used to effect variable reductions and exact solutions. The simplest time-dependent solution is shown to be stable, and relevant to a closed system.
}\par\vspace{7mm}}

\section{Introduction}

We consider the Lie group analysis of the class of radiant plasma energy equations of form
\begin{equation}\label{EqGAW}
u_t=\left[G(x)A(u)u_x\right]_x+W(t);\quad t\in(0,t_1),\quad x\in [0,x_1].
\end{equation}
As usual, $x$ and $t$ are respectively space and time coordinates.
$G$, $A$ and $W$ are arbitrary sufficiently smooth (analytic) functions of their variables.
Dependent variable $u$ represents energy density for which the main contribution is radiant energy,
often approximated by the equilibrium expression $\frac{4}{c}\epsilon\sigma T^4$, where $T$ is the absolute temperature,
$\sigma$ is the Stefan--Boltzmann constant, $\epsilon$ is the emissivity and c is the speed of light (e.g. \cite{Drake}).
$G(x)$ is mass density and $G(x)A(u)$ is the energy diffusivity. In general, $A$ may depend on temperature, and therefore on $u$,
so a linear model is inappropriate when the temperature is highly variable, due to laser heating.
In the setting of a plasma laboratory \cite{Lindl1998}, $W$ is the power of a laser used to heat the plasma.
For convenience, it will be assumed hereafter that a unit of length has been chosen so that $x_1=1$.

\medskip

In general, in realistic simulations the energy transport equation must be solved numerically.
However, some insight may be gained from special cases of (\ref{EqGAW}) that may be reduced to ordinary differential equations,
to algebraic equations or even to exact solutions. Therefore it is advantageous to have a full classification of models that allow symmetry reductions.
A classical Lie point symmetry classification has been done in \cite{Broadbridge&Lavrentiev&Williams1999} for the case of uniform density $G$.
However, it is a key feature of inertial confinement experiments that the laser blast causes some local compression.
Therefore we allow density to vary in space.
After completing the classical point symmetry classification, we consider local conservation laws and associated potential symmetries,
as well as some nonclassical symmetries. Finally, we construct some exact solutions.

We exclude some details of the cases $A=\const$ (linear equations) and $W(t)=\const$ (a complete classification of this case can be found
in the series of articles~\cite{Vaneeva&Johnpillai&Popovych&Sophocleous}).

\section{On group classification}

We start our investigation of symmetry properties of class~\eqref{EqGAW} by finding its group of usual equivalence transformations
and classification of Lie point symmetries.
Let us note that Lie symmetries of different subclasses of class~\eqref{EqGAW} have been investigated by many authors.
Thus, e.g., linear equations have been classified by Lie~\cite{Lie1881}
in his classification of linear second-order PDEs with two independent variables.
(See also a modern treatment of this subject in~\cite{Ovsiannikov1982}.)
Classification of nonlinear equations with $G(x)=1$ and $W(t)=0$ can be found in~\cite{Ovsiannikov1959}.
In~\cite{Popovych&Ivanova2004NVCDCEs} one can find, in particular, the group classification of~\eqref{EqGAW} with $W(t)=0$.
The series~\cite{Vaneeva&Johnpillai&Popovych&Sophocleous} contains as a particular case
the complete group classification of the subclass $W(t)=1$.
Symmetries of time-dependent equations~\eqref{EqGAW} with $G(x)=1$ have been found in~\cite{Broadbridge&Lavrentiev&Williams1999}.

First of all we look for the (point) equivalence group of class~\eqref{EqGAW}, i.e., transformations of dependent and independent
variables that move any equation from class~\eqref{EqGAW} to an equation of the same class.
Following the results of~\cite{Kingston&Sophocleous1998,Popovych&Ivanova2004NVCDCEs} for equivalence  transformations of parabolic equations,
without loss of generality we search for equivalence transformations of the form
\[
\tilde t=\tilde t(t), \quad \tilde x=\tilde x(t,x), \quad \tilde u=m(t,x)u+n(t,x).
\]
Substituting the above change of variables to~\eqref{EqGAW}, requiring the transformed equation to be
$\tilde u_{\tilde t}=(\tilde G(\tilde x)\tilde A(\tilde u)\tilde u_{\tilde x})_{\tilde x}+\tilde W(\tilde t)$,
we get an equation that is polynomial with respect to the derivatives of~$u$. Setting to zero its coefficients with respect to unconstrained variables
we get a system of overdetermined partial differential equations, general solution of which leads to the following theorem.

\begin{theorem}
Point equivalence group~$G^\sim$ of class~\eqref{EqGAW} consists of scaling and translation transformations of dependent and independent variables 
\begin{gather}\nonumber
\tilde t=\varepsilon_4t+\varepsilon_1 ,\quad
\tilde x=\varepsilon_5x+\varepsilon_2 ,\quad
\tilde u=\varepsilon_6u+\varepsilon_3 ,\\
\tilde A=\varepsilon_7 A ,\quad
\tilde G=\varepsilon_4^{-1}\varepsilon_5^2\varepsilon_7^{-1}G ,\quad
\tilde W=\varepsilon_4^{-1}\varepsilon_6W , \label{EquivTransf}
\end{gather}
where $\varepsilon_i=\const$, $\varepsilon_4\ldots\varepsilon_7\ne0$.
\end{theorem}

Below we perform the group classification of class~\eqref{EqGAW} up to the transformations~\eqref{EquivTransf}.

In the framework of the classical Lie--Ovsiannikov approach we search for an infinitesimal operator of Lie point symmetry of form
\[
X=\tau(t,x,u)\p_t+\xi(t,x,u)\p_x+\eta(t,x,u)\p_u.
\]
Splitting with respect to unconstrained variables the equation obtained from the infinitesimal invariance condition
(application of the second prolongation of~$X$ to~\eqref{EqGAW} on the manifold of~\eqref{EqGAW} is zero)
we get the following system for coefficients of the symmetry operator~$X$:
\begin{gather*}
\tau_x=\tau_u=\xi_u=\eta_{uu}=0,\\
GA(2\xi_x-\tau_t)- G_xA\xi- GA_u\eta=0,\\
\eta_t- W_t\tau-GA\eta_{xx}-G_xA\eta_x+W(\eta_u-\tau_t)=0,\\
GA_u(2\xi_x-\eta_u-\tau_t)-GA\eta_{uu}- GA_{uu}\eta- G_xA_u\xi=0,\\
GA(\xi_{xx}+\xi_x-2\eta_{xu})-G_xA\tau_t-\xi_t-G_xA_u\eta -G_{xx}A\xi -2 GA_u\eta_x=0.
\end{gather*}
An immediate consequence of these equations is $\eta_{uu}=0$. Then, from the rest of the equations we obtain
that either $\eta=0$, or, up to the equivalence group~\eqref{EquivTransf}, $A=u^m$ or $A=e^u$.
Solving the rest of the system of classifying equations separately in these three cases we get the following result.

\begin{theorem}
Allowing for point equivalence transformations~\eqref{EquivTransf},
equivalence classes of~\eqref{EqGAW} (with additional condition $W_tA_u\ne0$) that admit non-trivial Lie symmetry algebras,
are represented by the canonical forms in Tables~\ref{TableLieSymCaseAForAll}--\ref{TableLieSymCaseAum}.
\end{theorem}
In the tables, the universal quantifier denotes an arbitrary function.
\begin{center}
\footnotesize\setcounter{tbn}{0}\renewcommand{\arraystretch}{1.1}
Table \refstepcounter{tabu}\label{TableLieSymCaseAForAll}\thetabu. Case $\forall A$ \\[1ex]
\begin{tabular}{|l|c|c|l|c|}
\hline
N\vspacebefore & $G(x)$ & $W(t)$ & \hfill  $A^{\rm max}\hfill$ \\[0.5ex]
\hline
%
\refstepcounter{tbn}\label{caseAforallGxaWt-1}\thetbn.\vspacebefore &  $x^a$ &  $1/t$
& $\langle (a-2)t\p_t-x\p_x \rangle$     \\[0.5ex]
\refstepcounter{tbn}\label{caseAforallGexWt-1}\thetbn.\vspacebefore &  $e^{x}$ &  $1/t$
& $\langle t\p_t-\p_x \rangle$     \\[0.5ex]
\refstepcounter{tbn}\label{caseAforallGx2W}\thetbn.\vspacebefore &  $x^2$ &  $\forall$
& $\langle x\p_x \rangle$     \\[0.5ex]
\refstepcounter{tbn}\label{caseAforallG1W}\thetbn.\vspacebefore &  $1$ &  $\forall$
& $\langle \p_x \rangle$     \\[0.5ex]
\refstepcounter{tbn}\label{caseAforallG1Wt-1}\thetbn.\vspacebefore &  $1$ &  $1/t$
& $\langle \p_x, 2t\p_t+x\p_x \rangle$     \\[0.5ex]
\hline
\end{tabular}
\end{center}

Additionally to this there exist extra symmetries in the following cases

\begin{center}
\footnotesize\setcounter{tbn}{0}\renewcommand{\arraystretch}{1.1}
Table \refstepcounter{tabu}\label{TableLieSymCaseAexp}\thetabu. Case $A=e^{u}$ \\[1ex]
\begin{tabular}{|l|c|c|l|c|}
\hline
N\vspacebefore & $G(x)$ & $W(t)$ & \hfill  $A^{\rm max}\hfill$ \\[0.5ex]
\hline
\refstepcounter{tbn}\label{caseAexpGW}\thetbn.\vspacebefore &  $\forall$ &  $\forall$
& $\langle e^{-\int Wdt}(\p_t+W\p_u),\,$ \\[0.5ex]
  &&& $ e^{-\int Wdt}\Big(\int e^{\int Wdt}dt \p_t+(W\int e^{\int Wdt}dt-e^{-\int Wdt})\p_u\Big) \rangle$     \\[0.5ex]
\refstepcounter{tbn}\label{caseAexpGg1W}\thetbn.\vspacebefore &   $\frac1{Z''}$ &  $\forall$
& $\mathcal{A}_1$     \\[0.5ex]
\refstepcounter{tbn}\label{caseAexpGgW}\thetbn.\vspacebefore  &  $1$ &  $\forall$
& $\langle e^{-\int Wdt}(\p_t+W\p_u),\,\p_x,\, x\p_x+2\p_u,\,$ \\[0.5ex]
  &&& $ e^{-\int Wdt}\Big(\int e^{\int Wdt}dt \p_t+(W\int e^{\int Wdt}dt-e^{-\int Wdt})\p_u\Big) \rangle$     \\[0.5ex]
\hline
\end{tabular}
\end{center}
{\footnotesize
Here $Z=Z(x)$ is an arbitrary solution of ordinary differential equation
\[
-\frac{Z'''}{Z''}=\frac{3(mZ''+n)-a}{mZ+nx+p},
\]
$m$, $n$, $p$, $a$ are arbitrary constants, $\mathcal{A}_1$ is three-dimensional Lie algebra whose operators look like
$X=\tau\p_t+\xi\p_x+\eta\p_u$ with
$\xi=mZ+nx+p$, $\eta=-mZ_x-n+A-\tau_t$ and $\tau$ being a solution of equation
$\tau''+W\tau'+W'\tau=0$, that is
$$\tau=e^{-\int Wdt} \left[\int^t e^{\int^{t_1} ~W(t_2)dt_2}~dt_1+c_2\right].$$
}

\begin{center}
\footnotesize\setcounter{tbn}{0}\renewcommand{\arraystretch}{1.1}
Table \refstepcounter{tabu}\label{TableLieSymCaseAum}\thetabu. Case $A=u^m$ \\[1ex]
\begin{tabular}{|l|c|l|c|}
\hline
N\vspacebefore & $G(x)$ & $W(t)$ & \hfill  $A^{\rm max}\hfill$\\[0.5ex]
\hline
\refstepcounter{tbn}\label{caseAumGWt-(m+1)m}\thetbn.\vspacebefore &  $\forall$ &  $t^{-(m+1)/m}$
& $\langle mt\p_t-u\p_u \rangle$     \\[0.5ex]
\refstepcounter{tbn}\label{caseAumGxkWtn}\thetbn.\vspacebefore  &  $gx^k$ &  $t^n$
& $\langle (k-2)t\p_t-(mn+m+1)x\p_x+(k-2)(n+1)u\p_u \rangle$     \\[0.5ex]
\refstepcounter{tbn}\label{caseAumGxkWexpt}\thetbn.\vspacebefore  &  $gx^k$ &  $we^{t}$
& $\langle (k-2)\p_t-mx\p_x+(k-2)u\p_u \rangle$     \\[0.5ex]
\refstepcounter{tbn}\label{caseAumGexpxWtn}\thetbn.\vspacebefore  &  $ge^{x}$ &  $t^n$
& $\langle t\p_t-(mn+m+1)\p_x+(n+1)u\p_u \rangle$     \\[0.5ex]
\refstepcounter{tbn}\label{caseAumGexpxWexpt}\thetbn.\vspacebefore  &  $ge^{x}$ &  $we^{t}$
& $\langle \p_t-m\p_x+u\p_u \rangle$     \\[0.5ex]
\refstepcounter{tbn}\label{caseAumGx2Wt-(m+1)m}\thetbn.\vspacebefore  &  $gx^2$ &  $t^{-(m+1)/m}$
& $\langle x\p_x,\, mt\p_t-u\p_u \rangle$     \\[0.5ex]
\refstepcounter{tbn}\label{caseAumG1Wtn}\thetbn.\vspacebefore  &  $g$ &  $t^{n}$
& $\langle \p_x,\, 2t\p_t+(mn+m+1)x\p_x+2(n+1)u\p_u \rangle$     \\[0.5ex]
\refstepcounter{tbn}\label{caseAumG1Wexpt}\thetbn.\vspacebefore  &  $g$ &  $we^{t}$
& $\langle \p_x,\,2\p_t+mx\p_x+2u\p_u \rangle$     \\[0.5ex]
\hline
\end{tabular}
\end{center}
{\footnotesize
Here $g=\pm1$, $w=\pm1$.
In Cases~1 and~6 we require $m\ne-1$, to avoid case $W=\const$.

}

\section{Reductions to ordinary differential equations}

Lie symmetry analysis is one of very few available systematic tools used to construct solutions of (systems of) nonlinear partial differential equations.
Although, in contrast to the case of ordinary differential equations, it does not provide us with the general solution of a system,
in many cases it allows to construct wide classes of solutions being invariant with respect to symmetry transformations.
Roughly speaking, the main theorem on invariant solutions of differential equations~\cite{Ovsiannikov1982,Olver1986} states
that all solutions invariant with respect to an $r$-parametric solvable group of symmetries of the given $n$-dimensional system
of partial differential equations
(with some additional restrictions on the Lie algebras of infinitesimal invariance transformations)
can be obtained by solving a system of differential equations with $n-r$ independent variables.
In particular, if $r=n-1$, invariant solutions can be constructed via solving a system of ordinary differential equations.
Note, that reductions with respect to $n$-dimensional subgroups lead to algebraic equations.

In this paper we consider the case of a single equation with two independent variables.
Thus, to construct its solutions being invariant with respect to $1$-dimensional Lie algebra
spanned by infinitesimal symmetry generator of form  $v=\tau\p_t+\xi\p_x+\eta\p_u$,
we need to solve the following equation (sometimes called the invariant surface condition)
\[
\tau u_t+\xi u_x=\eta.
\]
Solution of this equation (its first integral)
gives an expression (Ansatz) for the invariants that can be used to define one new independent variable and the new dependent variable.
Substituting this in the initial equation we reduce it to an ordinary differential equation.
Any solution of the reduced equation yields an invariant solution of the initial equation.

This reduction procedure works only if the symmetry algebra satisfies the property of transversality.
For more details see~\cite{Ovsiannikov1982,Olver1986}.
Otherwise (in case of \emph{systems of} partial differential equations) one may try to look for so-called \emph{partially invariant solutions}.

As we discussed above, in general any subgroup of a (not necessarily point) Lie symmetry group corresponds to a class of invariant solutions
of (a system of) differential equations.
In  most cases there exists an infinite number of such subgroups.
This makes it practically impossible to construct directly all invariant solutions of the system even if its complete symmetry group is known.
An effective systematic way to classify such solutions
is to find a so-called ``optimal system'' of invariant solutions,
from which one can reconstruct all possible invariant solutions
by application of symmetry transformations to the solutions from the optimal system.

The background for construction of optimal systems of solutions of differential equations is given by the following statement~\cite{Ovsiannikov1982}:
Any two conjugate subgroups of a Lie symmetry group of a system of differential equations correspond to systems of reduced equations
that are related by a conjugacy transformation in the Lie symmetry group of the system
acting on the solutions invariant with respect to each subgroup.
Thus, all invariant solutions for a given system
can be constructed by selecting a subgroup in each conjugacy class of all symmetry subgroups.
Such a selection is called an optimal set (or optimal system) of subgroups of a Lie symmetry group.
Then we construct solutions being invariant with respect to the listed subgroups of symmetry transformations.
Action of the complete symmetry group on the above solutions give us all possible invariant solutions of the initial system.

A set of subalgebras of the Lie symmetry algebra corresponding to the optimal system of subgroups (called also ``optimal system of subalgebras'') consists of
subalgebras being inequivalent up to the actions of adjoint representation of the Lie symmetry group on its Lie algebra.
Ovsiannikov~\cite{Ovsiannikov1982} proposed an effective algorithm of construction of such optimal systems
(see also a nice simpler explanation and examples for optimal systems of one-dimensional subalgebras in~\cite{Olver1986}).
It is useful to note that  optimal systems of subalgebras of low-dimensional Lie algebras (dimension less than or equal to 4)
are classified in~\cite{Patera&Winternitz1977}.
Using these results we can easily write down all inequivalent subalgebras of  Lie symmetry algebras of
equations from class~\eqref{EqGAW} with power nonlinearity.

All equations presented in Table~3 possess 1-dimensional (Cases 1--5) or 2-dimensional (Cases 6--8) Lie symmetry algebras.

In Case~6 the invariance algebra is Abelian and its optimal system of subalgebras is
$\langle x\p_x \rangle$, $\langle mt\p_t-u\p_u+\varepsilon x\p_x\rangle$ ($\varepsilon=0,\pm1$), $\langle x\p_x,\, mt\p_t-u\p_u\rangle$.

Symmetry algebras in Cases 7--8 are non-Abelian with optimal systems of subalgebras
$\langle e_1 \rangle$, $\langle e_2 \rangle$, $\langle e_1, e_2\rangle$, where $e_1$, $e_2$ are the basis generators
of the corresponding algebras.
Even though the symmetry algebras in Cases 7--8 are non-Abelian, they are solvable so they still lead to successive
reductions.

In Table~4 we list all essential Lie reductions of equations from Table~3
(in each case the value of parameter functions for equations~\eqref{EqGAW} coincide with those from Table~3).
More precisely, we perform reductions of the corresponding equations with respect to the subalgebras from the adduced optimal systems.\\
The non-stationary reductions of equations from Cases 7 and 8 are partial cases of reductions of Cases 2 and 3 correspondingly.\\
Reductions with respect to subalgebras containing $x\p_x$ trivially lead to the $x$-independent solutions.

\pagebreak

\begin{center}
\renewcommand{\arraystretch}{1.2}
Table~4. Reduced ODEs for~\eqref{EqGAW}. \\
\footnotesize
\begin{tabular}{|l|c|c|l|}
\hline \vspacebefore
N & Ansatz $\tilde u=$& $\omega$ &\hfill {Reduced ODE\hfill} \\
\hline
1& $\varphi(\omega)t^{-1/m}$ & $x$ & $m(G\varphi^m\varphi')'+\varphi+m=0$\\
2& $\varphi(\omega)t^{n+1}$ & $xt^{\frac{mn+m+1}{k-2}}$ & $g(\omega^k\varphi^m\varphi')'+1-(n+1)\varphi-\frac{mn+m+1}{k-2}\omega\varphi'=0$\\
3& $\varphi(\omega)e^t$ & $xe^{mt/(k-2)}$ & $\varphi+\frac{m}{k-2}\omega\varphi'=g(\omega^k\varphi^m\varphi')'+w$\\
4& $\varphi(\omega)t^{n+1}$ & $x+(mn+m+1)\ln t$ & $(n+1)\varphi+(mn+m+1)\varphi'=g(e^{\omega}\varphi^m\varphi')'+1$\\
5& $\varphi(\omega)e^t$ & $x+mt$ & $m\varphi'+\varphi=g(e^{\omega}\varphi^m\varphi')'+w$\\
6& $\varphi(\omega)t^{-1/m}$ & $xt^{-\varepsilon/m}$ & $g(\omega^2\varphi^m\varphi')'+1+\frac1m\varphi+\frac{\varepsilon}{m}\omega\varphi'=0$\\
\hline
\end{tabular}
\end{center}

\section{Conservation laws and potential symmetries}

Roughly speaking~\cite{Olver1986} a {\em conservation law} of a system of
partial differential equations  $\mathcal{L}(x,u_{(r)})=0$ can be understood as a divergence expression
${\rm div}\,F=0$ which vanishes for all solutions of this system.
Here $x=(x_1,\ldots,x_n),$ $u=(u^1,\ldots,u^m).$
$F=(F^1,\ldots,F^n),$ where $F^i=F^i(x,u_{(r)}),$ is a conserved vector of this conservation law,
$u_{(r)}$ is the set of all partial derivatives of function $u$ with respect to $x$
of order not greater than~$r$, function $u$ itself is considered as the derivative of zero order.
The {\em order} of the conserved vector $F$ is the maximal order of derivatives that explicitly appear in $F$.

A conserved vector $F$ is called {\em trivial} if
$F^i=\hat F^i+\check F^i$, $i=\overline{1,n}$,
where $\hat F^i$ and $\check F^i$ are, likewise~$F^i$, functions of $x$ and derivatives of $u$,
$\hat F^i\equiv0|_{\mathcal L}$,
and the $n$-tuple $\check F=(\check F^1,\ldots,\check F^n)$
is a null divergence (i.e. its divergence vanishes identically).

A conservation law is called trivial, if its conserved vector is trivial.
Two conservation laws are equivalent, if their difference is a trivial conservation law.
Conservation laws are called linearly dependent if there exists a linear combination of them which is a trivial conservation law.

The above explanation is useful for the first intuitive illustration of notion of conservation laws.
However, to obtain complete understanding and to be able to describe correctly all possible conservation laws of (a system of) differential equations,
we should introduce a more rigorous definition of conservation laws
(see, e.g.,~\cite{Popovych&Ivanova2004ConsLawsLanl} for more details and examples).
Namely, for any system~$\mathcal{L}$ of differential equations the set~$\CV(\mathcal{L})$ of its conserved vectors is a linear space,
and the subset~$\CV_0(\mathcal{L})$ of trivial conserved vectors is a linear subspace in~$\CV(\mathcal{L})$.
The factor space~$\CL(\mathcal{L})=\CV(\mathcal{L})/\CV_0(\mathcal{L})$
coincides with the set of equivalence classes of~$\CV(\mathcal{L})$ with respect to the equivalence relation of conserved vectors.

\begin{definition}\label{DefinitionOfConsLaws}
The elements of the factor space~$\CL(\mathcal{L})$ are called {\em conservation laws} of the system~$\mathcal{L}$,
and~$\CL(\mathcal{L})$ is called {\em the space of conservation laws} of~$\mathcal{L}$.
\end{definition}

Thus, description of the conservation laws of a system~$\mathcal{L}$
is reduced to finding~$\CL(\mathcal{L})$
(more precisely, to construction of its basis if $\dim \CL(\mathcal{L})<\infty$ or a system of
generatrices if $\dim \CL(\mathcal{L})=\infty$).
The elements of~$\CV(\mathcal{L})$ from the same equivalence class define
conserved vectors of the same conservation law.
This allows us to identify the elements from~$\CL(\mathcal{L})$ with their representatives in~$\CV(\mathcal{L})$.
The {\em order of the conservation law} is called the minimum of the orders of the conserved vectors corresponding to this conservation law.
Linear (in)dependence of conservation laws is understood as linear (in)dependence of them as elements of~$\CL(\mathcal{L})$.
Thus, conservation laws of a system~$\mathcal{L}$ are called {\em linearly dependent} if there exists a linear combination of their representatives,
which is a trivial conserved vector.

\begin{note*}
If a (local) transformation of variables connects two systems of differential equations, then, by the action of this transformation, a conservation law of
the first of these systems is transformed into a conservation law of the second system,
i.e. the equivalence transformation establishes a one-to-one correspondence between conservation laws of these systems.
So, similarly to the problem of symmetry group classification,
 we can consider a problem of classification of conservation laws of a class of (systems of) differential equations with respect to its equivalence group.
(See~\cite{Popovych&Ivanova2004ConsLawsLanl} for more details, rigorous definitions and proofs.)
\end{note*}

Since equation~\eqref{EqGAW} is second-order evolutionary $(1+1)$-dimensional, then without loss of generality~\cite{Popovych&Ivanova2004ConsLawsLanl}
we can search for its conservation laws in form
\begin{equation}\label{EqConsLawGen}
D_tT(t,x,u)+D_xX(t,x,u,u_x)=0.
\end{equation}
Now, we expand the total derivatives in the above expression on the solution manifold of~\eqref{EqGAW}, take into account its differential consequences and
decompose the obtained expression with respect to the derivatives of~$u$. Solution of the resulting system of partial differential equations
gives the following theorem.

\begin{theorem}\label{TheoremLocalCL}
Any (nonlinear) equation from class~\eqref{EqGAW} admits exactly two linearly independent conservation laws
with densities and fluxes of form
\begin{gather}\textstyle
T= u-\int Wdt,\qquad X=-GAu_x ,\\ \textstyle
T=(u-\int Wdt)\cdot\int\frac{dx}G ,\qquad X=-GAu_x\cdot\int\frac{dx}G +\int Adu .
\end{gather}
\end{theorem}

\begin{note}
It  follows from existence of two linearly independent local conservation laws, that equation~\eqref{EqGAW} is equivalent
up to contact transformations to an equation of form $u_t=(F(t,x,u,u_x))_{xx}$
(see results of~\cite{Popovych&Samoilenko2008} for the second order $(1+1)$-dimensional evolution equations).
\end{note}

In the framework of group analysis of differential equations,
one of the natural applications of conservation laws is construction of non-local (potential) symmetries.
A system of differential equations may admit such symmetries when at least one of its equations
(or a differential consequence) can be written in a conserved form or, in other words,
the system possesses a non-trivial conservation law.
After introducing potentials from the conservation law as additional dependent variables,
we obtain a new (potential) system of differential equations.
Any local symmetry transformation of the obtained system induces a symmetry of the initial system.
If transformations of some of the ``non-potential'' local variables
explicitly depend on non-local potentials, this symmetry
is called a non-local (potential) symmetry of the initial system,
otherwise it projects into point symmetries of the initial system.
For more details about potential symmetries and their applications we refer the reader to~\cite{Bluman&Kumei1989}.

It follows from Theorem~\ref{TheoremLocalCL} that for equations from class~\eqref{EqGAW} there exist exactly two potential systems constructed
with the found conservation laws:
\begin{gather}\label{SysPotSys1}\textstyle
v_x=u-\int Wdt,\qquad v_t=GAu_x,\\ \textstyle
z_x=(u-\int Wdt)\cdot\int\frac{dx}G,\qquad z_t=GAu_x\cdot \int\frac{dx}G-\int Adu.
\label{SysPotSys2}
\end{gather}
(Here $v=v(t,x)$ and $z=z(t,x)$ are potential variables.)

It is known~\cite{Popovych&Ivanova2004ConsLawsLanl} that the equivalence group for a class of
systems of equations or the symmetry group for a single system can be naturally prolonged to potential variables.
One can use these prolonged equivalence groups for classification of potential symmetries of the given system.
In view of this statement we classify potential symmetries of class~\eqref{EqGAW}
up to the (trivial natural) prolongation of group~\eqref{EquivTransf} to the potential variables.

In the present work we reduce ourselves to classification of the potential symmetries of class~\eqref{EqGAW}
that arise from potential system~\eqref{SysPotSys1}.

Finding all Lie symmetries of the potential system~\eqref{SysPotSys1} we prove the following result.
\begin{theorem}
All inequivalent equations from class~\eqref{EqGAW} ($W(t)\ne0$) having potential symmetries that arise from system~\eqref{SysPotSys1}
are exhausted by the following ones:
\begin{itemize}
  \item
  $A(u)=u^{-2}$, $G(x)=1$, $W(t)=S'(t)$, $S(t)$ is a solution of the differential equation $S''(S^2+a_2)=a_1(S')^2$ ($a_2=0,\pm1$):\\
   $A_{\rm pot}=\langle \p_x, \p_v, a_1\frac{S'}{S''}\p_t+(v-\frac{a_1}2x)\p_x+[-u^2+(a_1+2S)u+a_1\frac{(S')^2}{S''}-a_1S-a_2-S^2]\p_u
   +(a_2x-\frac{a_1}2v)\p_v\rangle$.

  \item
  $A(u)=u^{-2}$, $G(x)=1$, $W(t)=w$: $A_{\rm pot}=\langle \p_x, \p_v, \p_t-wx\p_v, 2t\p_t-x\p_x+2u\p_u+v\p_v, wt^2\p_t+v\p_x-(u-2wt)u\p_u\rangle$.
\end{itemize}
(Together with the values of the arbitrary elements $A$, $G$ and $W$ for each case we adduce the potential symmetry algebras.)
\end{theorem}
We comment that the above second-order differential equation for $S(t)$, can be integrated twice to achieve a transcendental equation.
\begin{note}
As (to the best of our knowledge) there exists no systematic investigation of the potential symmetries of equations of form~\eqref{EqGAW}
with $W(t)\ne0$, in the above theorem for completeness we adduced also the constant-coefficient case.
\end{note}

Classification of the potential conservation laws and potential symmetries corresponding to other potential systems is a subject of a sequel paper.

  \section{Exact solutions}
  \subsection{The  case with no restriction except $A=e^u$}

From Table 2, this case has two independent additional symmetries. For an equation with this level of generality
(two free functions $G$ and $W$) to have an extra symmetry, it must be transformable to a simpler equation for which the symmetry is very simple.
Let $\mu=\exp(u)$.
  \begin{equation}
  \mu_t=\mu[G(x)\mu_x]_x+W(t)\mu~.
  \end{equation}
Let $\mu(x,t)=\Theta (x,t) exp(\int^t W(s)ds).$ Then
\[
e^{-\int W(t)dt}\Theta_t=[G(x)\Theta_x]_x,
\]
or more simply,
\begin{equation}
\label{simply}
\Theta_\tau=\Theta[G(x)\Theta_x]_x~,
\end{equation}
where
\[\tau=\int^t e^{\int^t W(s)ds}dt\,.
\]
Unlike the case of general energy diffusivity $A(u)$,
this equation is invariant under $\tau$-translation and therefore has a pseudo-steady state solution $\Theta=\Theta(x)$ satisfying
\[
G(x)\Theta_x=c,
\]
with $c$ an arbitrary constant.
From ~\eqref{EqGAW}, the heat flux is $-G(x)\mu_x$ which in the case of this solution, equates to a time dependent flux that is uniform in space,
taking the value $-c\,\exp(\int W(t)dt)$.  The solution for $\mu(x,t)$ takes the form of multiplicative separation of variables,
\begin{equation}
\mu=e^{\int W(t)dt}\left[c \int \frac{dx}{G(x)}+c_2\right],
\end{equation}
with $c_2$ arbitrary. That is equivalent to a  solution for $u(x,t)$ in the form of an additive separation of variables,
\begin{equation}
u=\int W(t)dt+\log\left(\int \frac{dx}{G(x)}~+c_4\right)+c_3,
\end{equation}
with $c_4$ and $c_3$ arbitrary constants.

\medskip

Equation \ref{simply} also has a scaling symmetry $\Theta\partial /\partial \Theta -\tau \partial /\partial \tau$.
Under this symmetry, invariant solutions have the form
$\Theta \tau=F(x)$, which is equivalent to
\[
u=-\log \tau + \log F(x)+\int^tW(s)ds,
\]
again an additive separated solution.

\subsection{A case with quadratic mass density}

Consider Table 3, Case 6. $A(u)=u^m, ~G(x)=x^2, ~W(t)=t^{-1-1/m}$.

\medskip

The two-dimensional symmetry algebra is generated by
$ax\partial_x+bmt\partial_t-bu\partial_u~;~~(a,b\in\Re).$
With $b\ne 0$, invariant solutions are of the form
\[
u=t^{-1/m}f(\phi)\,;\quad \phi=xt^{-a/(bm)},
\]
where
\begin{equation}
\label{reduction3.6} \frac{-1}{m}f(\phi)+\frac{-a}{bm}\phi f'(\phi)=[\phi^2f^mf'(\phi)]'+1.
\end{equation}
When $b=a$, this can be directly integrated to first order,
\[
f'(\phi)=\frac{-1}{m}\phi^{-1}f^{1-m}-\phi^{-1}f^{-m}-c_4\phi^{-2}f^{-m},
\]
with $c_4$ arbitrary.
This is an exactly solvable Bernoulli equation (as well as being a Riccati equation), only in the case $m=-1$.
This represents a heat diffusion coefficient that decreases with temperature. Then $g=1/f$ satisfies a first-order linear equation.
The two-parameter focussing solution is
\begin{equation}
u=t\,f(xt);\quad f(\phi)=\frac{c_5}{-\phi+c_6\phi e^{c_5/\phi}}.
\end{equation}
This equivalence class of equations  may also represent an increasing diffusion coefficient $A=(u_\infty-u)^{-1}$.
Here, $u_{\infty}$ is an upper bound to the energy density.
Then define $v=u_\infty-u$ that satisfies
\[v_t=\left(G(x)v^{-1}v_x\right)_x-1.
\]
This has the same symmetry reduction as above except that the constant source term 1 in (\ref{reduction3.6}) must be replaced by $-1$.
If $g=1/f$, then $g$ satisfies a linear first-order equation, leading to the two-parameter focussing solution
\begin{equation}
u=u_{\infty}-t\,f(xt);\quad f(\phi)=\frac{1}{-1+(c_6/\phi) e^{-c_6/\phi}E_i(c_6/\phi)+(c_5/\phi)e^{-c_6/\phi}}.
\end{equation}

\subsection{Spatially uniform solution}

The $x$-independent solution $u=Y(t)$ is available in the general case of ~\eqref{EqGAW}, not just in the special cases uncovered by classical Lie symmetry analysis. This is because equations of form~\eqref{EqGAW}
are $Q$-conditionally (nonclassically) invariant with respect to the operator $\p_x$.
In many practical cases of heated radiant plasma in a container with boundaries at $x=-1,1$,
the initial condition will be  symmetric and $u_x$ will remain zero at $x=0$.
In an ideal thermodynamically closed system, the container is perfectly insulated and hence there will be zero heat flux at $x=1$, implying that the Neumann boundary condition $u_x=0$ will apply also at that boundary. The total energy content is
\begin{gather*}
\nonumber
\int_0^1 u(x,t)dx=\bar u(t), \mbox{ and}\\
\frac{d\bar u}{dt}=\int_0^1 u_t(x,t)dx
=\int_0^1 \partial_x [G(x)A(u)u_x]+W(t)dx=W(t)
\end{gather*}
implying for any solution $u(x,t)$,
\[
\bar u(t)=\bar u(0)+\int_0^t W(s)ds=Y(t).
\]

Hence, in a closed system for which zero-flux Neumann boundary conditions must be applied, the spatially uniform solution is the time dependent
mean value of any spatially variable solution.
In the circumstance of a closed system, the diffusion process is expected to evolve any solution asymptotically towards
the spatially uniform mean value $u=Y(t)$. This solution is stable, as can be seen from linear stability analysis.
Assume $u=Y(t)+v(x,t)$ is a solution of ~\eqref{EqGAW}, $v$ being a small perturbation with
\[
v^2+v_x^2=O(\varepsilon^2).
\]
Then to order $\varepsilon$,
\begin{gather*}
\nonumber
v_t=\partial_x[G(x)A(Y(t))v_x]\\
\implies v_\tau=\partial_x[G(x)v_x],
\end{gather*}
where $\tau=\int_0^tA(Y(s))ds$.
The equation for $v(x,\tau)$ is non-denerate provided $G$ has a minimum value $G_{min}>0$, in which case it is a dissipative parabolic equation.
Since both $u$ and $Y$ satisfy Neumann boundary conditions, so must  $v_x=0$ at $x=0,1.$
The mean value of $u$ is fully accounted for in $Y(t)$, so the mean value of $v$ must be zero.
Since it has zero gradient at the boundaries, the solution for $v$ dissipates to a function with zero gradient and zero mean value, that is $v\to 0$.

\subsection{An integrable model}

The second equivalence class given in Theorem 4, represented by
\[
u_t=\partial_x\left[u^{-2}u_x\right]+1
\]
is in fact an integrable case. Some other members of this equivalence class were used in~\cite{Broadbridge&Rogers1999}.

In the case of  energy transport in a plasma, the minor modification $u=v-b$ gives a model that has diffusivity increasing with energy density
\[
v_t=\partial_x\left[\frac{1}{(b-v)^2}v_x\right]+1.
\]
However the energy density $v$ must then be restricted to be less than $b$.

Let $u=\phi_x$ for some potential $\phi$. Then it is sufficient that
\[
\phi_t=\frac{1}{\phi_x^2}\phi_{xx}+x.
\]
Using the identity $\phi_t t_x x_\phi=-1$, the hodograph transformation results in Burgers' equation
\[
x_t=x_{\phi\phi}-xx_{\phi}.
\]
This gives the possibility of constructing some relevant solutions for  plasma energy.

If we also allow a small amount of energy leakage at $x=1$, then we impose a constant-flux $J=R$ boundary condition at $x=1$.
With a constant sink, rather than a constant source term and with the boundary labels $x=0,1$ interchanged,
this problem has already been transformed in~\cite{Broadbridge&Rogers1999} to a linear boundary value problem.
An identical treatment applies with a constant source.
In principle, but with some difficulty, arbitrary initial conditions with $v(x,t)<b$ may be treated in this way.

\section{Conclusion}
When the mass density is allowed to be non-uniform, the equation for radiant energy transport in a laser-heated plasma, has a richer Lie point symmetry algebra, an additional non-obvious local conservation law and associated potential symmetries. These have been used to effect variable reductions and to construct some special solutions in closed form.  When the nonlinear energy diffusivity depends exponentially on energy density, there is a special symmetry allowing reduction by separation of variables, independent of the form of variable density and laser intensity. When the density is a squared linear function of  coordinate $x$ and the energy diffusivity and the laser intensity are particular powers of energy density and time respectively, the nonlinear partial differential equation has a symmetry reduction to an ordinary differential equation that may, in some cases, be solved in terms of elementary functions and exponential integrals. These solutions may be indicative of energy distributions when in the future, the mass density can be better predicted and controlled.
The potential symmetry classification shows that any model with a genuine potential symmetry is equivalent to a spatially homogeneous model but with three possible types of time-dependent laser intensity. The special case of constant laser intensity and inverse-square nonlinear diffusivity, not only allows a second potential symmetry, but is in fact  an integrable nonlinear model that may be transformed to Burgers' equation. A nonclassical, rather than classical symmetry classification is required to pick up the trivial spatially uniform but time-dependent solution that applies to the general form of Equation (\ref{EqGAW}). In a thermodynamically closed system, this simple solution is stable and it is in fact simply the spatial arithmetic mean of the more general spatially dependent solution.

\end{document}